\documentclass[english,reprint,floatfix]{revtex4-2}
\usepackage{mathptmx}

\usepackage{courier}

\usepackage[T1]{fontenc}
\usepackage[latin9]{inputenc}
\usepackage{geometry}
\usepackage{color}
\usepackage{babel}
\usepackage{float}
\usepackage{amsmath}
\usepackage{graphicx}
\usepackage[unicode=true,
 bookmarks=true,bookmarksnumbered=true,bookmarksopen=true,bookmarksopenlevel=1,
 breaklinks=true,pdfborder={0 0 1},backref=false,colorlinks=true]
 {hyperref}

\hypersetup{pdftitle={High temperature embedding sDFT},
 pdfauthor={Minh Nguyen},
 linkcolor=black,citecolor=black,urlcolor=blue,filecolor=blue,pdfpagelayout=OneColumn,pdfnewwindow=true,pdfstartview=XYZ,plainpages=false}

\makeatletter

\@ifundefined{date}{}{\date{}}
\usepackage{babel}
\usepackage{chemformula}
\usepackage{braket}
\makeatother

\begin{document}
\title{Time-Dependent Density Functional Theory with the Orthogonal Projector
Augmented Wave Method}
\author{Minh Nguyen}
\affiliation{Department of Chemistry and Biochemistry, University of California, Los Angeles,
Los Angeles, CA, 90095, USA}
\author{Tim Duong}
\affiliation{Department of Chemistry and Biochemistry, University of California, Los Angeles,
Los Angeles, CA, 90095, USA}
\author{Daniel Neuhauser}
\affiliation{Department of Chemistry and Biochemistry, University of California, Los Angeles,
Los Angeles, CA, 90095, USA}

\begin{abstract}
The projector augmented wave (PAW) method of Bl\"{o}chl linearly maps smooth pseudo wavefunctions to the highly oscillatory all-electron DFT orbitals.
Compared to norm-conserving pseudopotentials (NCPP), PAW has the advantage of lower kinetic energy cutoffs and larger grid spacings at the cost of having to solve for non-orthogonal wavefunctions.
We earlier developed orthogonal PAW (OPAW) to allow the use of PAW when orthogonal wavefunctions are required. 
In OPAW, the pseudo wavefunctions are transformed through the efficient application of powers of the PAW overlap operator with essentially no extra cost compared to NCPP methods. 
Previously, we applied OPAW to DFT. 
Here, we take the first step to make OPAW viable for post-DFT methods by implementing it in real-time time-dependent (TD) DFT. 
Using fourth-order Runge-Kutta for the time-propagation, we compare calculations of absorption spectra for various organic and biological molecules and show that very large grid spacings are sufficient, 0.6-0.8$~$Bohr in OPAW-TDDFT rather than the 0.4-0.5$~$Bohr used in traditional NCPP-TDDFT calculations. 
This reduces the memory and propagation costs by up to a factor of 5. 
Our method would be directly applicable to any post-DFT methods that require time-dependent propagations such as GW and BSE.
\end{abstract}
\maketitle

\section{Introduction}

In plane-wave basis set electronic structure calculations, convergence is determined by the kinetic energy cutoff parameter.
Plane waves are well suited for sufficiently smooth wavefunctions but require a large basis set to accurately represent highly oscillatory atomic core states. 
One popular method to circumvent this issue is the use of norm-conserving pseudopotentials (NCPP) in which the effects of the inert core states are replaced with an effective pseudopotential, leading to smooth pseudo valance wavefunctions that can be easily represented in a plane-wave basis \citep{troullier1991efficient}.
However, the simplicity of NCPP comes at the price of high computational costs in treating first-row elements, transition metals, and rare earth elements \citep{hamann1979norm}.

The projector augmented wave (PAW) method, first proposed by Bl\"{o}chl, yields smoother pseudo wavefunctions than those of NCPP by relaxing norm conservation \citep{blochl1994projector}. 
These smoother wavefunctions have a lower kinetic energy cutoff and, therefore, can use a larger grid spacing than in NCPP.
The key to PAW is a linear transformation that maps the smoothed pseudo wavefunctions to the highly oscillatory all-electron wavefunctions. 
This transformation enables calculation of wavefunction-dependent properties such as hyperfine parameters, core-level spectra, electric-field gradients, and NMR chemical shifts \citep{pickard2001all}.

This linear mapping results in non-orthogonality of the pseudo wavefunctions that satisfy instead a generalized eigenvalue problem which complicates the use of PAW in electronic structure methods that rely on orthogonal wavefunctions such as stochastic density functional theory (DFT) \citep{baer2013self} or stochastic GW \citep{neuhauser2014breaking}.
We recently solved this non-orthogonality problem through the combination of two techniques that were discovered earlier \citep{li2020real}. 
The first is the efficient application of powers of the PAW overlap operator to generate an orthogonal Hamiltonian and wavefunctions that are norm-preserving \cite{ljungberg2011implementation,li2020real}.
The second is the Ono-Hirose transformation which yields accurate overlaps of coarse-grid wavefunctions with the localized dense-grid atomic projector functions \cite{ono1999timesaving}.
With these two ingredients, our resulting orthogonal-PAW (OPAW) method was demonstrated with the Chebyshev-filtered subspace iteration DFT approach, successfully reproducing PAW band gaps from the ABINIT software \citep{li2020real,zhou2014chebyshev}.

Many post-DFT methods, such as our stochastic GW approximation \citep{neuhauser2014breaking} and stochastic Bethe Salpeter equation methodologies \citep{bradbury2022bethe}, are easier to implement with orthogonal time-dependent wavefunctions.
Here, we take the next step in developing OPAW, making it viable for time-dependent (TD) DFT using real-time propagation. 
Compared to linear-response methods, real-time propagation can calculate electronic responses to any arbitrary external stimuli, allowing for study of non-linear-response phenomena such as high-harmonic generation and exciton dynamics in photovoltaic devices \citep{pauletti2021role,petrone2014charge}.

Our combined OPAW-TDDFT approach directly uses orthogonal wavefunctions and the Ono-Hirose method to yield an efficient real-time TD propagation method which, as we will show, works well even with very large grid spacing and increased time-steps. 
Note that other implementations of PAW have earlier been used for time-dependent propagation \citep{walter2008time,enkovaara2010electronic,akimov2013pyxaid}.

The OPAW-TDDFT method is demonstrated here with absorption spectra calculations for various organic- and chromophore-based systems.
Section \ref{sec:Methodology} reviews the OPAW theory and discusses its implementation in TDDFT.
In section \ref{sec:Results}, we present and analyze absorption spectra calculations using OPAW-TDDFT against those calculated with NCPP-TDDFT, and time-step analysis follows in \ref{sec:computation_time}.
Concluding remarks follow in section \ref{sec:conclusion}.

\section{\label{sec:Methodology}Theory}

\subsection{\label{subsec:OPAW-DFT}Orthogonal Projector Augmented Wave Method}

In PAW, the all-electron (AE) wavefunctions, $\psi_{n}$, are built from the pseudo (PS) wavefunctions, $\tilde{\psi}_{n}$, using the linear map:
\begin{equation}
 \ket{\psi_{n}}=\hat{T}\ket{\tilde{\psi}_{n}}=\ket{\tilde{\psi}_{n}}+\sum_{a,i}\big(\ket{\phi_{i}^{(a)}}-\ket{\tilde{\phi}_{i}^{(a)}}\big)\braket{p_{i}^{(a)}|\tilde{\psi}_{n}},
\end{equation}
where $a$ indexes the atoms and $i$ stands for a combination of angular, magnetic, and principal quantum numbers to label each partial wave channel associated with an atom. 
The AE ($\phi_{i}^{(a)})$ and PS ($\tilde{\phi}_{i}^{(a)}$) partial wave channels equal each other outside a spherical augmentation region around each atom. 
The atomic projector functions, $p_{i}^{(a)}$, are localized within the augmentation regions and are dual to the PS partial waves there (i.e. $\sum_{i}\ket{\tilde{\phi}_{i}^{(a)}}\bra{p_{i}^{(a)}}\simeq1$). 

Since the PS wavefunctions are not orthogonal, they fulfill a generalized eigenproblem,
\begin{equation}
 \tilde{H}\tilde{\psi}_{i}=\varepsilon_{i}\hat{S}\tilde{\psi}_{i},
\end{equation}
where 
\begin{equation}
 \hat{S}=\hat{T}^{\dagger}\hat{T}=\mathbf{I}+\sum_{ij,a}\ket{p_{i}^{(a)}}s_{ij}^{(a)}\bra{p_{i}^{(a)}}
\end{equation}
is an overlap matrix with $s_{ij}^{(a)}\equiv\braket{\phi_{i}^{(a)}|\phi_{j}^{(a)}}-\braket{\tilde{\phi}_{i}^{(a)}|\tilde{\phi}_{j}^{(a)}}$,
and
\begin{equation}
 \tilde{H}=-\frac{\nabla^{2}}{2}+v_{KS}(r)+\sum_{ij,a}\ket{p_{i}^{(a)}}D_{ij}^{(a)}\bra{p_{i}^{(a)}}.
\end{equation}
Full details of the effective potential, $v_{KS}(r)$, and the non-local term coefficients ($D_{ij}^{(a)}$) can be found in Refs. \citep{blochl1994projector,torrent2008implementation}.

To make OPAW, we rotate the PS wavefunctions to make orthogonal PS wavefunctions,
\begin{equation}
 %\bar{\psi}_{i}=\hat{S}^{1/2}\tilde{\psi}_{i}\label{eq:wf_bar},
 \psi'_{i}=\hat{S}^{1/2}\tilde{\psi}_{i}\label{eq:wf_bar},
\end{equation}
resulting in 
\begin{equation}
 %\bar{H}\bar{\psi}_{i}=\varepsilon_{i}\bar{\psi}_{i}\label{eq: orthogonal_schro},
 H'\psi'_{i}=\varepsilon_{i}\psi'_{i}\label{eq: orthogonal_schro},
\end{equation}
where $H'=\hat{S}^{-1/2}\tilde{H}\hat{S}^{-1/2}$ is the OPAW Hamiltonian. 
To efficiently approximate $\hat{S}^{n}$, where $n$ is any real number, one first assumes that the augmentation regions of different atoms do not overlap so that the projector functions can be separately rotated around each atom \citep{li2020real}. 
With this assumption, $\hat{S}$ is readily transformed into 
\begin{equation}
 \hat{S}=\mathbf{I}+\sum_{i,a}\ket{\eta_{i}^{(a)}}o_{i}^{(a)}\bra{\eta_{i}^{(a)}},
\end{equation}
where the rotated projectors, ${\eta_{i}^{(a)}}$, are orthogonal so that $\braket{\eta_{i}^{(a)}|\eta_{j}^{(a')}}=\delta_{ij}\delta_{a,a'}$ and $o^{(a)}_i$ are derived from transformations applied to $s^{(a)}_{ij}$.
We verified earlier \citep{li2020real} that the projector locality assumption yields accurate band gaps, and will show below how it is also sufficient for time-dependent properties.
Any power of $\hat{S}$ is then easily expressed as
\begin{equation}
 \hat{S}^{n}=\mathbf{I}+\sum_{i,a}\ket{\eta_{i}^{(a)}}[(1+o_{i}^{(a)})^{-\frac{1}{2}}-1]\bra{\eta_{i}^{(a)}}.
\end{equation}

The second key to the efficacy of OPAW is the use of the Ono-Hirose method to efficiently calculate the overlap of the denser grid projector functions with the coarser grid wavefunctions in any application of $\hat{S}^{n}$ and $\hat{S}$ \citep{ono1999timesaving}.
In the method, a smoothing matrix that connects the dense-grid and the coarse-grid is constructed using spline interpolation in the $x$, $y$, and $z$ directions which when applied onto the dense-grid projector functions generates coarse-grid projector functions.
These coarse-grid projector functions are then used in calculating coarse-grid overlaps.
Further details of the transformation of the projectors and the application of the Ono-Hirose method are given in Ref. \citep{li2020real}.

\subsection{\label{subsec: OPAW-TDDFT} Time-Dependent Density Functional Theory
with the Orthogonal Projector Augmented Wave Method}

One common use of TDDFT is the calculation of absorption spectra of materials and molecules. 
The absorption cross section in the linear response regime is formally
\begin{equation}
 \sigma(\omega)=\frac{4\pi}{c}\omega\int drdr'\delta\tilde{v}(r,\omega)\tilde{\chi}(r,r',\omega)\delta\tilde{v}(r',\omega),\label{eq: absorption_cross_section_1}
\end{equation}
where $\delta\tilde{v}(r,\omega)$ is the perturbing external potential and $\tilde{\chi}(r,r',\omega)$ is the susceptibility function for $\omega>0$. 
For absorption, $\delta v$ can be a dipole potential pulse polarized in the, say, $x$ direction,
\begin{equation}
 \delta v(r,t)=x\delta(t)\gamma, \label{delta v}
\end{equation}
where $\gamma$ is a small perturbation strength.
With this potential, Eq. (\ref{eq: absorption_cross_section_1}) is obtained through the dipole-dipole correlation function $
 d_{xx}(t)=\frac{1}{\gamma}\int d{r}\,x\Delta n_{x}(r,t)\
$ where 
\begin{equation}
 \Delta n_{x}(r,t)=\int dr'\tilde{\chi}(r,r',t)v(r')=\frac{1}{\gamma}\big(n^{\gamma}(r,t)-n^{\gamma=0}(r,t)\big)
 \label{eq: induced_charge_density}
\end{equation}
is the induced charged density and 
\begin{equation}
 n^{\gamma}(r,t)=\sum_{i}f_{i}|\Psi^{\prime\gamma}_{i}(r,t)|^{2}
\end{equation}
are the perturbed and unperturbed densities made from the time-dependent OPAW wavefunctions, labeled by $\gamma$, and detailed later.

The absorption spectrum is then given by an average of the Fourier transform of the dipole-dipole correlation function in the $x$, $y$, and $z$ directions,
\begin{equation}
 \sigma(\omega)=\frac{4\pi\omega}{3c}\int_{0}^{\infty}dt
 \left(d_{xx}(t)+d_{yy}(t)+d_{zz}(t)\right) e^{i\omega t}.
 \label{eq: sigma}
\end{equation}

The time-dependent Schr\"odinger equation in the PAW framework has the form
\begin{equation}
 i\hat{S}\frac{\partial}{\partial t}\tilde{\Psi}_i^\gamma(r,t)=\tilde{H}\tilde{\Psi}_i^\gamma(r,t),
\end{equation}
where $\tilde{\Psi}_i^\gamma(r,t)$ is a time-dependent PS wavefunction evolving under the PAW Hamiltonian. 
With the transformations in Eqs. (\ref{eq:wf_bar}) and (\ref{eq: orthogonal_schro}), this equation in the OPAW framework becomes
\begin{equation}
 i\frac{\partial}{\partial t}\Psi_i^{\prime\gamma}(r,t)=H'(t)\Psi_i^{\prime\gamma}(r,t),
 \label{eq: opaw td-schrodinger eq}
\end{equation}
where $\Psi_i^{\prime\gamma}(r,t)$ is the time-dependent orthogonalized PS wavefunction. 
The initial condition of the OPAW wavefunction for an $x$ polarized potential in Eq. (\ref{delta v}), is 
\begin{equation}
 \Psi_i^{\prime\gamma}(r,t=0)=e^{-i \gamma x' } \psi_i(r)
\end{equation}
where $x'=\hat{S}^{-1/2} x \hat{S}^{-1/2}$. 
In practice we find that we could replace $x'$ by $x$ without any significant change in the results.

\begin{figure}[htbp]
 \includegraphics[width= 0.85\columnwidth]{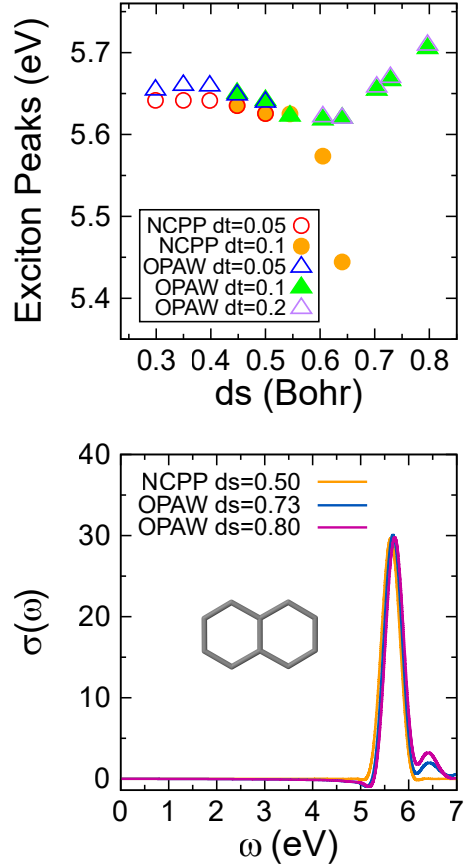}
 \caption{The top panel shows naphthalene exciton peaks vs. grid spacing with various time-steps (in a.u.) using OPAW and NCPP.
The bottom panel shows naphthalene absorption spectra calculated using OPAW for very large grid spacings (in Bohr) and a large time-step 0.2$~$a.u. 
 The results match an NCPP simulation at a lower grid spacing and a smaller time-step $0.1$ a.u.}
 \label{fig. naphthalene convergence}
\end{figure}

For simplicity, we use fourth-order Runge-Kutta to approximate the time-propagation of the OPAW wavefunctions in Eq. (\ref{eq: induced_charge_density}). 
Compared with other typically used approaches, e.g., the split-operator method, the Runge-Kutta approach is simple to implement as it does not require exponentiation of non-local terms in the Hamiltonian; for further details see Sec. \ref{sec:computation_time}.

\section{\label{sec:Results}Results}
We demonstrate below OPAW-TDDFT and show that it allows the use of much larger grid spacings than NCPP-TDDFT. 
For simplicity, we abbreviate the two methods as OPAW and NCPP, omitting the TDDFT label.

For both methods, the ground-state wavefunctions were calculated with the Chebyshev-filtered subspace iteration method \citep{zhou2014chebyshev}. 
The DFT Hamiltonian used the LDA functional. 
The simulations were non-periodic and employed the Martyna-Tuckerman approach for the Coulombic interactions \citep{martyna1999reciprocal}.
For OPAW, we employed the recommended LDA-based atomic datasets from the ABINIT website \citep{jollet2014generation}, while for NCPP we used the Hamann form for the pseudopotentials \citep{hamann1979norm}. 
The simulation boxes for all calculations were chosen so that there was approximately 6 Bohr of padding from the edges of the system in the $x,y$, and $z$ directions.

\begin{figure}[htbp]
 \begin{centering}
 \includegraphics[width= 0.85\columnwidth]{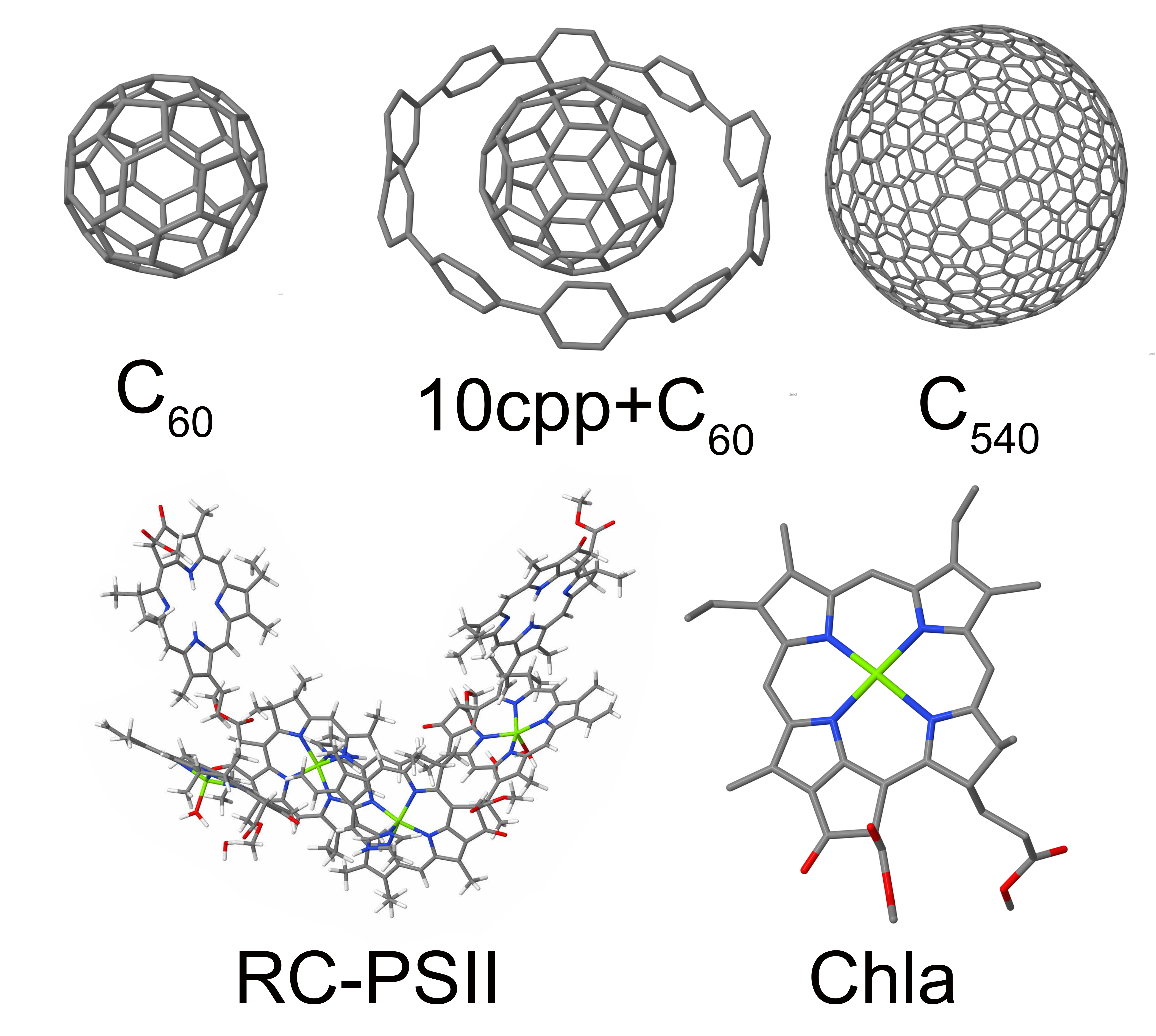} 
 \par\end{centering}
 \noindent \raggedleft{}
 \caption{Structures and abbreviations for all systems used in this paper.}
 \label{fig: 5 big mols}
\end{figure}

The time-dependent propagations for both OPAW and NCPP used the Runge-Kutta method, with time-steps that are either 0.05, 0.1 or 0.2 a.u. 
Specifically, most of the results shown here use the largest time-step for which the simulations are stable for a particular system and grid spacing.
Generally, OPAW enables the use of larger time steps than what is possible with NCPP, as detailed below.

Since for each molecule the box size is unchanged between runs, the grid spacings are generally slightly different in the $x$, $y$, and $z$ directions, and therefore we report the results against the grid spacings' geometrical averages, $ds\equiv\left({dx \, dy \, dz}\right)^{1/3}=dV^{1/3}$.

We first study the convergence with grid spacing using a small molecule (naphthalene). 
The simulations use a box size of $28\ {\rm {Bohr}}\times26\ {\rm {Bohr}}\times16\ {\rm {Bohr}}$ with $ds$ in the range of 0.3$~$Bohr to 0.8$~$Bohr.

\begin{figure}[htbp]
 \includegraphics[width= 0.85\columnwidth]{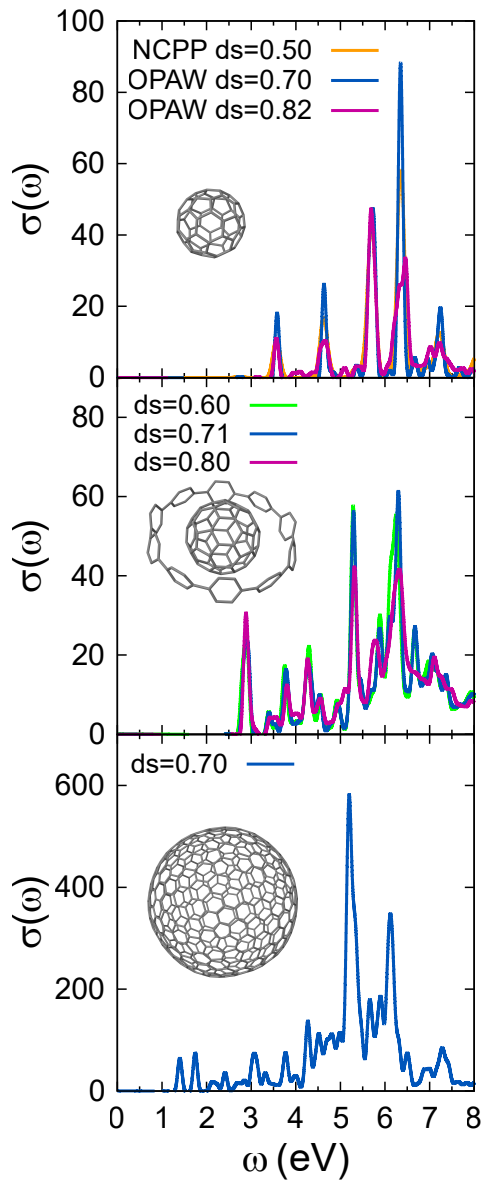}

 \caption{The top panel shows OPAW absorption spectra of fullerene for large grid spacings with a time-step $0.2$ a.u., which match the results from NCPP; the latter required a smaller grid spacing and a smaller time-step $0.1$ au. 
 The middle and bottom panels show the OPAW absorption spectra for 10CPP+C$_{60}$ and C$_{540}$ respectively for large grid spacings with a time-step $0.2$ a.u.}
 \label{fig: fullerene_spectra}
\end{figure}

\begin{figure}[htbp]
 \includegraphics[width= 0.85\columnwidth]{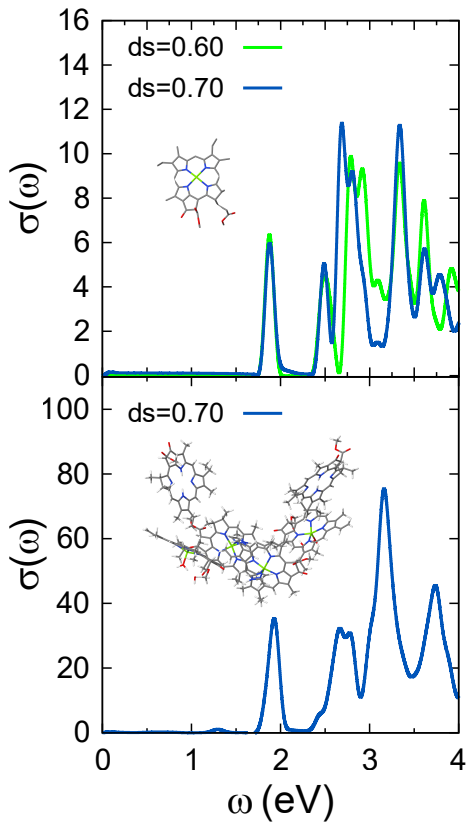}

 \caption{The top and bottom panels show the OPAW absorption spectra of Chla and RC-PSII respectively for large grid spacings with a time-step $0.1$ a.u.}
 \label{fig: chromophores spectra}
\end{figure}

Fig. \ref{fig. naphthalene convergence} shows the exciton peaks (i.e., first significant absorption peak) for several spectra calculated at different grid spacings and time-steps for NCPP and OPAW.
Compared to NCPP, OPAW shows excellent spectral peaks at much larger grid spacing. 
Specifically, note that even at $ds=0.8$ Bohr, the OPAW exciton peak is still acceptable, i.e., different from the low-$ds$ values by only \textasciitilde$0.05$ eV. 
We also show in Fig. \ref{fig. naphthalene convergence} that the spectra for OPAW at high $ds$ values are well converged compared to NCPP at lower $ds$ values.
 
Our convergence test with naphthalene demonstrates that grid spacings of 0.6-0.8 Bohr are sufficient to converge OPAW at a level that requires $ds=0.4$-$0.5$ Bohr for NCPP. 
These results correspond to computational savings in the 3D grid of up to a factor of 3-5.

Our next step is to demonstrate the power of the OPAW on a series of five larger systems, shown in Fig. \ref{fig: 5 big mols}. 
Included are C$_{60}$ fullerene, C$_{60}$ fullerene embedded inside a 10 para-substituted phenyl "nanohoop" cycloparaphenylene ring (10CPP+C$_{60}$) \citep{xu2018concave}, a C$_{540}$ fullerene, a Chlorophyll-a chromophore with a methyl acetate ligand in place of the phytyl chain (Chla), \citep{forster2022quasiparticle} and a hexameric reaction center of photosystem II consisting of six chromophores (RC-PSII) \citep{forster2022quasiparticle}. 
A total simulation time of 2000 a.u. was used to generate spectra for all five systems, which is sufficient to isolate specific peaks to better than 0.05 eV.

As shown in Fig. \ref{fig: fullerene_spectra}, OPAW generates comparable spectra to that of NCPP at much larger grid spacing for C$_{60}$. 
Similarly for the larger 10CPP+C$_{60}$ and C$_{540}$ systems, OPAW also calculates converged spectra at large grid spacing, around 0.7-0.8 Bohr.

Compared to the hydrocarbon-based systems, the Chla and RC PSII systems with OPAW require slightly smaller grid spacings (0.7 rather than 0.8 Bohr), but these are still quite large compared to those typical of NCPP. 
Although some of the peaks of the two OPAW spectra of Chla in Fig. \ref{fig: chromophores spectra} start to differ significantly for energies greater than 2.7 eV, the lower-energy peak positions match well.

\section{Computation Time\label{sec:computation_time}}

We now turn to a discussion of the time requirements and computational savings in OPAW.
We verified that the time to construct the Hamiltonian and perform Runge-Kutta propagation at every time step is similar between OPAW and NCPP given the same time-step and grid and that both scale linearly with system size and with the number of grid points (i.e., quadratically overall with system size).
Thus, for our largest system (RC-PSII) with 676 occupied orbitals and a grid with 822,848 points, the total wall time to perform a single OPAW-TDDFT calculation on RC-PSII, using a single modern node with 104 AMD-ROME cores, was almost 20 Wall hours, i.e., about 2,000 CPU core hours; the times for the smaller systems are correspondingly smaller.

When comparing OPAW and NCPP, one should note that NCPP could have used a non-Runge-Kutta approach, i.e., a symplectic (split-operator) propagation (a symplectic representation of OPAW is more complicated and was not pursued here).  
But while symplectic propagation is fast per time step, it necessitates a shorter time-step (for the split operator, we verified that for most systems the time-step can be at most 0.05 a.u. with NCPP). 
The costs per time-step are different in the two methods, and we find that a symplectic propagation typically requires half the cost of Runge-Kutta, due to the balance of several terms:
\begin{itemize}
\item In Runge-Kutta, the cost per time step includes four Hamiltonian operations and a single calculation and ``broadcasting'' of the time-dependent potential across all cores;
\item in symplectic methods, the costs include calculating the kinetic energy propagator which is the most expensive part, two local and non-local potentials, and the calculation and ``broadcasting'' of the time-dependent potential.
\end{itemize}

Thus, overall, the cost of NCPP with symplectic propagation with a time-step 0.05 a.u. is comparable to that of using Runge-Kutta with a time-step 0.1 a.u.  
Given that the Runge-Kutta time-step is 0.1 or 0.2 a.u. in our calculations, the lack of symplectic propagation is not a problem for OPAW.

\section{Conclusions\label{sec:conclusion}}
Our OPAW method has been implemented here in the TDDFT framework using fourth-order Runge-Kutta for the time propagation. 
We demonstrated that the convergence of OPAW with respect to the grid spacing and time-step is achieved at much larger values compared to that of NCPP, resulting in significant computational savings.

We also showed, using a series of hydrocarbons and chromophore based systems, that OPAW successfully produces converged spectra at large grid spacing and time-steps.
Future work will focus on implementing OPAW into our other post-DFT stochastic methods that require orthogonal time-dependent wavefunctions, namely stochastic \textit{GW} \citep{neuhauser2014breaking} and  Bethe-Salpeter Equation \citep{bradbury2022bethe,rabani2015time} methods.

\section*{Acknowledgments}
We are grateful to Nadine Bradbury and Tucker Allen for useful discussion. 
This work was supported by the U.S. Department of Energy, Office of Science, Office of Advanced Scientific Computing Research, Scientific Discovery through Advanced Computing (SciDAC) program, under Award No. DE-SC0022198. 
Computational resources were provide through ACCESS allocations Discover allocation PHY220143.

\bibliographystyle{unsrt}
\bibliography{opaw_tddft}

\end{document}